\documentclass[aip,reprint,floatfix]{revtex4-1}

\usepackage{savesym}

\usepackage{graphicx}
\usepackage{bm}
\savesymbol{setrefcountdefault}
\usepackage[bookmarks=false]{hyperref}

\begin{document}


\title{Magnetically guided Cesium interferometer for inertial sensing} 




\author{Lu Qi}
\email[]{qilu\_alex@buaa.edu.cn}
\affiliation{School of Instrumentation Science and Opto-Electronics Engineering, Beihang University, Beijing, 100191,China}

\author{Zhaohui Hu}
\affiliation{School of Instrumentation Science and Opto-Electronics Engineering, Beihang University, Beijing, 100191,China}

\author{Tristan Valenzuela}
\affiliation{RAL Space, STFC Rutherford Appleton Laboratory, Didcot, OX11 0QX, United Kingdom}

\author{Yuchi Zhang}
\affiliation{College of Physics and Electronics Engineering, Shanxi University, Taiyuan, 030006, China}

\author{Yueyang Zhai}
\affiliation{School of Instrumentation Science and Opto-Electronics Engineering, Beihang University, Beijing, 100191,China}

\author{Wei Quan}
\affiliation{School of Instrumentation Science and Opto-Electronics Engineering, Beihang University, Beijing, 100191,China}

\author{Nick Waltham}
\affiliation{RAL Space, STFC Rutherford Appleton Laboratory, Didcot, OX11 0QX, United Kingdom}

\author{Jiancheng Fang}
\email[]{fangjiancheng@buaa.edu.cn}
\affiliation{School of Instrumentation Science and Opto-Electronics Engineering, Beihang University, Beijing, 100191,China}


\date{\today}

\begin{abstract}
In this paper we demonstrate a magnetically guided Cesium (Cs) atom interferometer in the Talbot-Lau regime for inertial sensing with two interferometer schemes, Mach-Zehnder and Ramsey-Borde. The recoil frequency of the Cs atoms and the acceleration along the waveguide symmetry axis is measured. An acceleration measurement uncertainty of $7\times10^{-5}$ m/s$^{2}$ is achieved. We also realize an enclosed area of $0.018$ mm$^{2}$ for rotation measurement. As the first reported magnetically guided Cs atom interferometer, the system limitation and its advantages are discussed.
\end{abstract}


\maketitle 

Since its first demonstration in the 1990s\cite{one}, atom interferometry has developed rapidly and i s now widely used in applications such as fundamental physics and precision measurements. Among these applications, atom interferometers are found especially useful for inertial sensing because of their huge potential in sensitivity enhancement compared to classical sensors. Gravimeters\cite{two,three}, gravity gradiometers\cite{four,five} and rotation sensors\cite{six,seven,eight} are built with different designs. Some of these atom interferometer sensors in the laboratory are now competing with, or even surpassing the performance of the best instruments using conventional methods\cite{three,nine}. Depending on the environment in which the atoms are manipulated, these sensors can be divided into two categories: free-space sensors and wave-guided sensors. The free-space sensor is usually based on an atomic fountain scheme, which is widely used in gravity and gravity gradient measurements. In order to significantly increase the interrogation time to improve the sensitivity of the instruments, longer falling distance and larger system volume are required, which limits their practical application\cite{three}. However, sensors based on wave-guided atom interferometer provide a chance for the combination of both long interrogation time and a compact volume, especially in the measurements of rotation and horizontal acceleration\cite{ten,eleven}. Two kinds of wave-guides, optical\cite{ten} and magnetic\cite{eleven,twelve,thirteen}, have been used. A proof-of-principle experiment of an atom interferometer with horizontal optical guide for inertial sensing has been realized with a Rubidium (Rb) Bose-Einstein Condensate (BEC) and a sensitivity of $7\times10^{-4}$ m/s$^{2}$, the performance of which is limited by the interparticle interaction in the BEC and vibration noise\cite{ten}.
An atom gravimeter in a vertical optical guide was achieved with a small falling distance of 0.8 mm and a sensitivity of $2\times10^{-7}g$  over 300s integration. The contrast loss was attributed to the spatial quality of the guide laser beam\cite{PhysRevA.85.013639}. The study of magnetically guided atom interferometers with laser cooled atoms started earlier. The first inertial sensor has been demonstrated using magnetically guided Rb atoms, with precisely machined ferromagnetic foils\cite{eleven}. A resolution of 10 times the Earth’s rotation was achieved. Another magnetically guided Rb interferometer was realised with race track coils\cite{twelve}. The improvement on interrogation time is limited by the atom-guide interaction in both atom interferometers\cite{twentythree,twelve}. Other guide schemes with more homogeneous field and more compact size are being explored\cite{lesanovsky2007time-averaged,twentyfour}. The combination of an atom wave guide and atom interferometry is expected to both improve the sensitivity and compactness of the instrument. 

\begin{figure}
\includegraphics{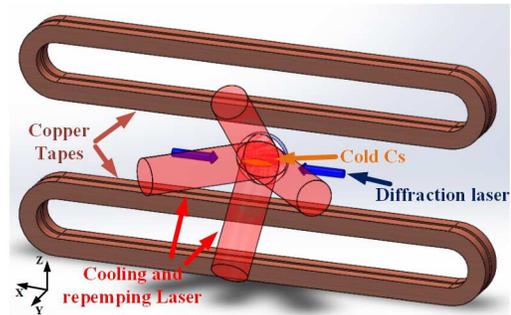}
\caption{\label{fig:sch}Schematic drawing of the experiment system}
\end{figure}

In this paper, we demonstrate a magnetically guided Cesium (Cs) atom interferometer for inertial sensing. The guided Cs atom interferometer is based on a dual-chamber vacuum system, which is detailed in previous work\cite{sixteen}. Briefly, a Cs atom beam is generated by a 2D+ magneto-optical trap (MOT) in the first chamber.
Passing through two differential tubes, the atom beam arrives at the second chamber with a beam flux of up to $2.5\times10^{10}$ atoms/s.
Atoms from the beam are then recaptured by a mirror MOT in the second chamber, with two cooling beams on the x-z plane, two cooling beams on y-z plane and their reflections respectively, as show in Fig.\ref{fig:sch}.
The angle between each beam and the z-axis is about 45 degrees. Each cooling beam contains 8 mW of cooling laser power  ($-2\Gamma$ detuning from $6^{2}s_{1/2}|F=4\rangle\to6^{2}p_{3/2}|F'=5\rangle$) and 2 mW  of repumping laser power  (resonant with $6^{2}s_{1/2}|F=3\rangle\to6^{2}p_{3/2}|F'=4\rangle$), with a Gaussian width of 10 mm.
The magnetic field for the MOT and magnetic guide is provided by the same two pairs of copper coils wound on two race-track aluminum cores.
These aluminum cores are mounted on a precision rotation stage (Thorlabs NR360), which provides precise angle position with respect to the y-axis. Each set of coil consists of N=30 turns of copper tapes (6.25 mm wide and 0.25 mm thick ) with a 0.05 mm thick Kapton tape on one side.
In the mirror MOT regime, a current of 10 A provides a radial gradient of 25 Gauss/cm in the area of the symmetry center, which is set to shorten the displacement during the atom loading.
The loading process is as follows: After a 900 ms loading time of the mirror MOT, the coil current begins to ramp from 10 A to 30 A over 60 ms, during which the MOT was compressed to match the mode of the magnetic guide.
At the beginning of the compression, the intensity of the cooling laser is attenuated to half of its original value; and the cooling frequency detuning ramps from $-2\Gamma$ to $-6\Gamma$. The cooling and repumping light is kept on to reduce the heating effect during the compression and the repumping light is switched off 4 ms before the cooling light, which helps to pump atoms into $6^{2}s_{1/2}|F=3\rangle$.
Then, an extra 20 ms waiting time is given before the interferometer operation. 
This enables the non-weak-field-seeking atoms to leave the beam area, removing pseudo contributions to signals.
The sequence parameters in the loading procedure are optimized by detecting the signal amplitude of a three-pulse atom interferometer with short interrogation time T (here we choose T=274 $\mu$s), the signal amplitude of which depends mostly on the number of atoms in the guide.
The number of atoms in the loading procedure is measured using an open-transition ($6^{2}s_{1/2}|F=3\rangle\to6^{2}p_{3/2}|F'=4\rangle$) absorption method\cite{seventeen}.A number of atoms of $1\times10^{9}$ is obtained
Finally, about $5\times10^{7}$ atoms are detected in the guide with a 85 $\mu$m Gaussian width, which indicates a 5\% efficiency of the procedure. The trap frequency in the radial direction is calculated to be $2\pi\times98$ Hz, indicating that the radial temperature of the Cs atoms is 34 $\mu$K. This procedure is featured with its simplicity in operation and shows a tolerance on the beam alignments experimentally. However, the efficiency could be further improved by polarizing the atom spin using optical pumping. 

During the interferometer manipulation, atoms are diffracted in the Kapitza-Dirac regime to form a Talbot-Lau interferometer\cite{eighteen}. Two off-resonant lasers, $E^{a}$ and $E^{b}$, (both 200 MHz blue-detuned from $6^{2}s_{1/2}|F=3\rangle\to6^{2}p_{3/2}|F'=4\rangle$) form the diffraction grating along x-axis, the symmetry axis of the waveguide, with 2 mm Gaussian width and a peak intensity of 127 mW/cm$^{2}$. The phase and amplitude information of the interference fringe is probed by monitoring the Bragg backscattered light from the $E^{a}$ into the $E^{b}$ mode. The information contained in the backscattered light is then extracted using the heterodyne technique\cite{eighteen}. 

\begin{figure}
\includegraphics{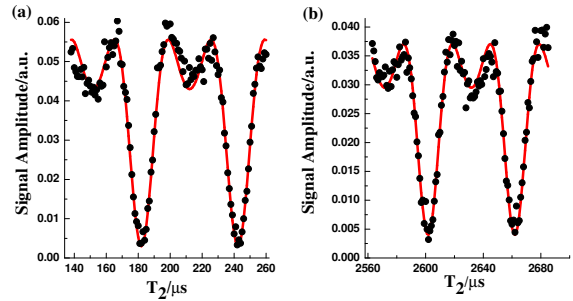}
\caption{\label{fig:omegaq}(Color online) Mean signal amplitude oscillation and data fit vs $T_{2}$. Data scanned in the period (a) between 138 $\mu$s to 260 $\mu$s and (b) between 2563 $\mu$s to 2685 $\mu$s. Experimental data(black dots) is fitted to a function of $f=f_{1} \cdot J_{2}(2\Theta_{2}sin(\omega_{q}T_{2}+f_{2}))+f_{0}$(red line).}
\end{figure}

Due to the relatively low diffraction efficiency of the diffraction pulse\cite{eighteen}, three-pulse and four-pulse regimes are most commonly used in the experiments. For the three-pulse Mach-Zehnder scheme (two diffraction pulses, one detection pulse and the time-set $\{T_{1}=0,T_{2},T_{3}=2T_{2}\}$, the fringe distribution ${\rho}_{3}(\bm{x},t)$ is simplified as:
\begin{equation}
\rho _3(\bm{x},t)\propto \Theta_1 \omega_q \Delta t J_2 (2\Theta _2\sin (\omega_q T_2))e^{i\mathbf{q}\cdot \mathbf{x}+i \phi_3(T_3)}
\end{equation}
where $\Theta_{i}$ represents the pulse area for the $i$th diffraction pulse, $\omega_{q}$ is the recoil frequency, $\Delta t=t-T_{3}$ indicates the time window before the fringe is washed out due to the thermal expansion, ${{\phi }_{3}}(T_{3})=\mathbf{q}\cdot \bm{a}{{T_{3}}^{2}}/4$ is the phase induced by the acceleration $\bm{a}$. The mean signal amplitude during the time window could be further simplified to $\langle\rho_{3}(t)\rangle_{\Delta t}\propto J_{2}(2\Theta_{2}sin(\omega_{q}T_{2}))$, which oscillates with the Cs recoil frequency. This oscillation is observed by scanning $T_{2}$. Two examples are plotted in Fig.\ref{fig:omegaq} with a 2 $\mu$s step. A fit of the experiment results with free parameters including $\Theta_{2}$ and $\omega_{q}$ showed a good agreement with the theoretical expectation.
The recoil frequency derived from the fit is $\omega_{q}=5.19\pm0.03$ MHz, compared to the theoretical value of $\hbar q^{2}/2m_{Cs}=5.197$ MHz. The pulse area is $\Theta_{2}=1.99\pm0.02$, compared to the theoretical value $\int\Omega_{DR}(\tau)d\tau=1.982$, where $\Omega_{DR}$ is the double photon Rabi frequency.
By minimizing the scan step and increasing the scan length, this method could be used to measure the fine structure constant $\alpha$\cite{PhysRevLett.101.230801}.

\begin{figure}
\includegraphics{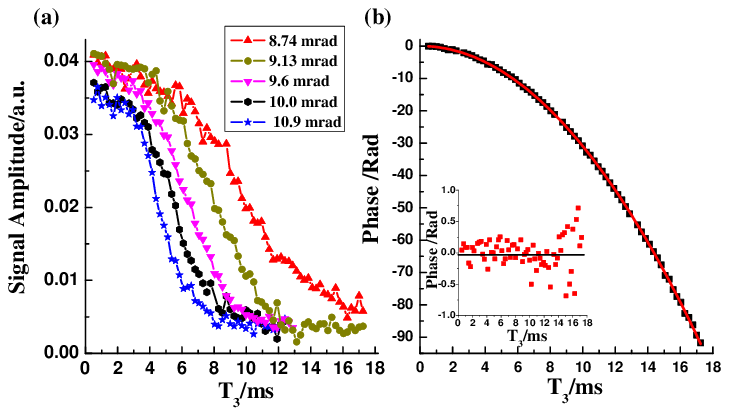}
\caption{\label{fig:3pulse}(Color online)(a)Signal amplitude decay of the three-pulse Mach-Zehnder interferometer with different inclination angles $\theta_{y}$ of the magnetic guide. The values of $\theta_{y}$ are extracted from the fit of the relative phase data. Here we assume gravity acceleration $g=9.8$ m/s$^{2}$. (b)Unwrapped phase data(black squares) and data fit(solid line) at the optimum incline angle $\theta_{y_{0}}=8.74$ mrad. The inset shows the fit residuals(red squares). The fit of $f=f_{1}\cdot T_{3}^{2}+f_{0}$ gives $f_{1}=-0.3157\pm0.0003$ and $f_{0}=0.01\pm0.05$, showing an acceleration uncertainty of 0.1mm/s$^{2}$}
\end{figure}

Due to the oscillation feature of the signal amplitude, the onset time of the second pulse was set to be $T_{2}=137+n\times121$ $\mu$s to keep the signal amplitude at the peak of the oscillation in the following experiments.
We now consider the signal amplitude and phase evolution of the three-pulse interferometer with a total interrogation time $T_{3}$. Fig.\ref{fig:3pulse} shows the results for the three-pulse interferometer with different inclination angles $\theta_{y}$ of the magnetic guide with respect to the y-axis.
For different values of $\theta_{y}$, atoms in the guide gain different accelerations due to the various projections of the gravity. 
This in turn impacts both the signal amplitude and the phase of the interferometer. 
By scanning $\theta_{y}$ with a precision level of sub-mrad, we fix the optimum value of $\theta_{y_{0}}=8.74$ mrad, in which the symmetry axis overlaps with the diffraction beam. At the optimum position, a maximum interrogation time of $\sim$18 ms is achieved.
This interrogation time is mainly limited by the curvature and roughness of the magnetic field along symmetry axis\cite{twentythree}.
As $\theta_{y}$ varies from the optimum value, the signal amplitude decays faster to the noise limit due to the strong decoherence mechanism from the projection of the in-homogeneous radial magnetic field.
A trend of smaller difference between two adjacent signal amplitude decay curves is observed as $\varepsilon$ increases, which is expected according to the prediction that the decoherence scales with $T_{3}^{2}$ at a given deviation angle $\varepsilon$\cite{twentythree}. The phase evolution and data fit as a function of total interrogation time $T_{3}$ at the optimum position is plotted in Fig.\ref{fig:3pulse}(b). A parabolic fit of $\phi(t)$ gives the acceleration $a$ along the waveguide symmetry axis over 70 runs. A diverging trend for phase residuals is also observed in the inset of Fig.\ref{fig:3pulse}(b), as the noise $\delta\phi(T_{3})=\sqrt{\langle[\phi(0)-2\phi(T_{3}/2)+\phi({T_{3}})]^{2}\rangle_{t}}$ becomes larger as $T_{3}$ increases.

\begin{figure}
\includegraphics{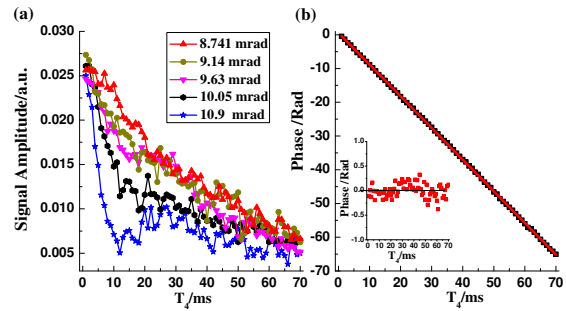}
\caption{\label{fig:4pulse}(Color online)(a)Signal amplitude decay of the four-pulse Ramsey-Borde interferometer with different inclination angles $\theta_{y}$ of the magnetic guide. The values of $\theta_{y}$ are extracted from the fit of the relative phase data. Here we suppose gravity acceleration $g=9.8$ m/s$^{2}$. (b)Unwrapped phase data(black squares) and data fit(solid line) at the optimum incline angle $\theta_{y_{0}}=8.741$ mrad. The inset shows the fit residuals(red squares). The fit of $f=f_{1}\cdot T_{4}+f_{0}$ gives $f_{1}=-0.9392\pm0.0008$ and $f_{0}=0.57\pm0.03$, showing an acceleration uncertainty of 0.07 mm/s$^{2}$}
\end{figure}

We now consider the four-pulse interferometer regime (three diffraction pulses, one detection pulse and the time set $\{T_{1}=0,T_{2},T_{3}, T_{4}\}$).
In comparison to the three-pulse regime, several degenerate loops exist with different phase contributions\cite{twenty}. One simple way to remove this degeneracy is to elongate the period between $T_{2}$ and $T_{3}$, which only results in a trapezoid loop and thus a Ramsey-borde type interferometer.
The fringe distribution is simplified as:
\begin{equation}
\rho_4(\bm{x},t)\propto \Theta_1 \omega_q \Delta t{J_1}^2(2\Theta \sin (\omega_q T_2))e^{i\mathbf{q}\cdot \mathbf{x}+i \phi_4(T_{2},T_{4})}
\end{equation}

where $\phi(T_{2},T_{4})=\bm{q}\cdot\bm{a}T_{2}(T_{4}-T_{2})$. Fig.\ref{fig:4pulse} shows the experiment results in the same positions as the three-pulse interferometer experiment with a fixed $T_{2}$ and thus a constant split distance $d=\hbar q T_{2}/m_{Cs}=5.24$ $\mu$m. The signal amplitudes in Fig.\ref{fig:4pulse}(a) decay with slower rates than those in the three-pulse interferometer at the same $\varepsilon$, which is due to the fixed split distance.
The phase evolution with the total interrogation time $T_{4}$ at the optimum angle in Fig.\ref{fig:4pulse}(b) fits nicely as expected. The acceleration is extracted from the phase over 70 runs.
The inset of Fig.\ref{fig:4pulse}(b) shows that noise level stays stable with $T_{4}$, compared to a divergence in the three-pusle phase fit. This is due to the fact that the long holding time between $T_{2}$ and $T_{3}$ makes the vibration noises uncorrelated for the splitting and combining processes; and the noise $\delta\phi(T_{4})=\sqrt{2\langle[\phi(0)-\phi(T_{2})]^{2}\rangle_{t}}$ is uncorrelated with $T_{4}$\cite{twentytwo}.

\begin{figure}
\includegraphics[width=8.5cm]{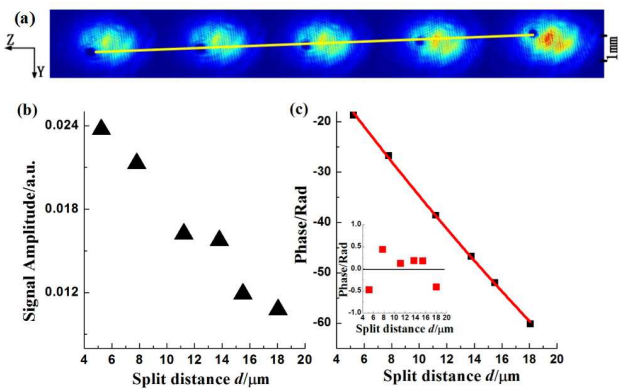}
\caption{\label{fig:move}(Color online) (a) Absorption image showing the guided atom motion in 20 ms, the yellow line is used to show the position of atoms. (b)Signal amplitude(solid triangles) of the four-pulse Ramsey-borde interferometer in the moving guide regime vs. the split distance. (c) Unwrapped phase data(black squares) and data fit(solid line) vs. the split distance. The inset shows the fit residuals(red squares). The fit of $f=f_{1}\cdot d^{2}-\hbar Q T_{4} f_{1}/m\cdot d+f_{0}$ gives $f_{1}=-0.0276\pm0.0003$ and $f_{0}=1.4\pm0.5$}.
\end{figure}

The slow decay rate of the Ramsey-Borde interferometer makes rotation measurements available. 
For the three-pulse interferometer, even at the optimum position, only about 18 ms of total interrogation time was achieved, which limits the guiding distance of the atoms. Therefore, it is experimentally difficult to achieve an area-enclosed interferometer using the three-pulse regime in our system. For the Ramsey-Borde interferometer, the $\sim$100 ms  long interrogation time supports a 2 mm distance with slow enough guiding speed, which prevents the atoms from being heated during the guiding operation.
In addition, the fixed distance difference helps to maintain momentum coherence between the two arms, which is avoided from being disturbed by the increasing curvature along the symmetry axis as the atoms move towards the edge of the copper tapes. Fig.\ref{fig:move} shows the experimental results of Ramsey-Borde interferometer in a 20 ms moving magnetic guide, during which the amplitude of the three-pulse interferometer vanishes.
During the guide process, the current in one pair of tapes was controlled from 30 A to 35 A with a sinusoidal waveform (phase from $-\pi/2$ to $\pi/2$); while the other pair was controlled from 30 A to 25 A  (phase from  to $\pi/2$ to $3\pi/2$).
The absorption images Fig.\ref{fig:move} (a) shows a calibrated guide distance of 1 mm; thus the average speed of the atoms is 50 mm/s.
The signal amplitude vs split distance is shown in Fig.\ref{fig:move}(b). The largest split distance achieved in the experiment is $d=18.06$ $\mu$m, the signal of which almost reaches the noise limit for phase detection. This split distance is short compared to that achieved in the three-pulse regime, which is mainly due to the increased longitudinal curvature outside of the symmetrical area. Unfortunately, we cannot input a constant rotation to the optical table due to technical constraints.
However, it is reasonable to induce a linear acceleration to evaluate the phase, for the detected acceleration $\bm{a}$ contains both the linear and Coriolis acceleration. As it is shown in Fig.\ref{fig:move}(c), the phase induced by gravity projection agrees with the predicted phase evolution.
In practice, it is feasible to align the magnetic guide normal to the gravity to erase the contribution from gravity. Guiding the atoms in opposite directions and differentiating the phase signal would erase the linear acceleration as a common mode signal. For $d=18.06$ $\mu$m, the enclosed area A is 0.018 mm$^{2}$, and a rotation rate of 1 mrad/s induces a phase shift of $0.24\pi$. 

To further increase the sensitivity of the interferometer, an improved guide, with more homogeneous field along the symmetry axis, is needed,in which the atoms separate further and thus a longer interrogation time can be achieved.
As the atoms are affected by the magnetic field along the symmetry axis, the magnetic field gradient causes an extra phase shift. 
This could be avoided by using an optical guide\cite{ten} or eliminated by operating in a reciprocal interfering loop\cite{eleven} in rotation measurement.
However, one advantage for Cs atom inteferometers is that the Cs atoms heavier mass results in a larger phase shift than the one obtained in a Rb atom interferometer in rotation measurements for the same magnetic guide. Consider the rotation phase:
\begin{eqnarray}
\phi_{r}=2m_{atom}\bm{\Omega}\cdot\bm{A}/\hbar\propto m_{atom}\Omega d_{c}l
\end{eqnarray}
where $d_{c}$ is the seperate distance between two arms and $l$ is the guide distance, both of which are mainly determined by the magnetic guide. For Cs atoms at $|F=3,m_{F}=-3>$ and Rb atoms at $|F=1,m_{F}=-1>$, the value of the magnetic field gradient to sustain either of them is the same. 
As a result, the rotation phase shift of Cs is $m_{Cs}/m_{Rb}\approx1.5$ times bigger than that of the magnetically guided Rb interferometer\cite{twelve}.

In summary, we report for the first time the realization of a magnetically guided Cs atom interferometer for inertial sensing.
An uncertainty of $7\times10^{-5}$ m/s$^{2}$ on horizontal acceleration measurements is achieved and an area of 0.018 mm$^{2}$ is enclosed for rotation measurements.
The current limitations on this interferometer system are revealed and the advantage of the greater mass of Cs in rotation measurement is discussed.
Improved guide configurations are required for longer interrogation times and larger guiding distance in the future. Our work could also be used as a reference to the multi-axis inertial measurements for practical use.  



%
%

%

\begin{acknowledgments}
This work is supported by National Natural Science Foundation of China(GrantNo. 61374210 and 61227902), the National Key R\&D Program of China(Grant No. 2016YFB0501600).
\end{acknowledgments}

\bibliography{Atominterferometer}

\end{document}